\documentclass[aps,prd,twocolumn,superscriptaddress,nofootinbib,10pt]{revtex4-1}
\usepackage[utf8]{inputenc}
\usepackage{booktabs}   
\usepackage{multirow}   
\usepackage{float}
\usepackage[colorlinks,linkcolor=blue,anchorcolor=blue,citecolor=blue,urlcolor=blue]{hyperref}
\usepackage{graphicx,subfigure}
\usepackage[figuresright]{rotating}
\usepackage{amsmath,amsfonts,amssymb,bm}
\usepackage{array,enumitem,multirow}
\usepackage{acronym}
\usepackage{makecell}
\newcommand{\be}{\begin{equation}}
\newcommand{\ee}{\end{equation}}
\newcommand{\bea}{\begin{eqnarray}}
\newcommand{\eea}{\end{eqnarray}}
\newcommand{\nn}{\nonumber}

\newcommand{\TQC}{MOE Key Laboratory of TianQin Mission, TianQin Research Center for Gravitational Physics \&  School of Physics and Astronomy, Frontiers Science Center for TianQin, Gravitational Wave Research Center of CNSA, Sun Yat-sen University (Zhuhai Campus), Zhuhai 519082, China.}

\newacro{GR}{General Relativity}
\newacro{GW}{Gravitational Wave}
\newacro{BGRS}{Beyond GR Signal}
\newacro{BMS}{Bondi-Sachs-Metzner}
\newacro{FRW}{Friedmann–Robertson–Walker}
\newacro{BBN}{Big Bang Nucleosynthesis}
\newacro{FBH}{Fluid Black Hole}
\newacro{MBHB}{Massive Black Hole Binary}
\newacroplural{MBHB}{Massive Black Hole Binaries}
\newacro{EMRI}{Extreme Mass Ratio Inspiral}
\newacro{IMRI}{Intermediate Mass Ratio Inspiral}
\newacro{SBHB}{Stellar-mass Black Bole Binary}
\newacroplural{SBHB}{Stellar-mass Black Bole Binaries}
\newacro{GCB}{Galactic Compact Binary}
\newacroplural{GCB}{Galactic Compact Binaries}
\newacro{SGWB}{Stochastic Gravitational Wave Background}
\acrodef{VB}{Verification Binary}
\acrodefplural{VB}{Verification Binaries}
\acrodef{EWPT}{Electroweak Phase Transition}
\newacro{MBH}{massive black hole}
\newacro{SBH}{stellar-mass black hole}
\newacro{PBH}{primordial black hole}
\newacro{iEMRI}{induced Extreme Mass Ratio Inspiral}
\newacro{QNM}{Quasi-Nomral Mode}
\newacro{ISCO}{Innermost Stable Circular Orbit}
\newacro{LVK}{the LIGO-Virgo-Kagra collaboration}
\newacro{CE}{Cosmic Explorer}
\newacro{ET}{Einstein Telescope}
\newacro{LISA}{Laser Interferometer Space Antenna}
\acrodef{PSD}{Power Spectral Density}
\newacro{SNR}{Signal-to-Noise Ratio}
\newacro{FIM}{Fisher Information Matrix}
\newacro{MCMC}{Markov Chain Monte Carlo}
\newacro{xKS}{extended Kerr-Schild}

\begin{document}

\title{Testing general relativity with binary black holes: a study on the sensitivity requirements for future space-based detectors}

\author{Tangchao Zhan}
\author{Changfu Shi}
\affiliation{\TQC}
\author{Shuo Sun}
\affiliation{School of Physical Science and Technology, Kunming University, Kunming 650214, China}
\author{Jianwei Mei}
\email{Corresponding author. Email: meijw@sysu.edu.cn}
\affiliation{\TQC}

\date{\today}

\begin{abstract}
We study the sensitivity required for a future space-based detector to search for beyond general relativity effect in gravitational wave detection.
To do this, we use the current design of TianQin, LISA, and $\mu$Ares as starting points, and study how their key noise parameters should be improved to adequately detect some target signals,
for which we choose a nonlinear ringdown mode, displacement memory, and a putative beyond general relativity signal, all from the merger of massive black hole binaries.
We find that the required improvements are strongly dependent on the choice of the target signals and the population model of massive black hole binaries, 
and $4-9$ orders of magnitude improvement will be needed in the most demanding detection scenarios.
\end{abstract}

\maketitle

\section{Introduction}
\label{sec:intro}

The breakthrough in \ac{GW} detection \cite{Abbott:2016blz} has given us the opportunity to probe the nature of gravity under conditions not possible before \cite{Gair:2012nm,Yunes:2013dva,Berti:2015itd,Yagi:2016jml,Barack:2018yly,Cardoso:2019rvt,Barausse:2020rsu,LISA:2022kgy,Luo:2025ewp,Berti:2025hly}.
However, so far no violation of \ac{GR} has been identified in the detected \ac{GW} signals \cite{LIGOScientific:2016lio,LIGOScientific:2019fpa,LIGOScientific:2020tif,LIGOScientific:2021sio,KAGRA:2025oiz,LIGOScientific:2025obp}.
This then leads to an intriguing question: 
If no \ac{BGRS} is found in any of the current \ac{GW} detection missions, to what extent should one continue to improve the detection sensitivities?
An answer to this question can help us better understand 
the prospect of searching for \ac{BGRS} simply by improving the sensitivity of a detector and
the level of technological improvement that will be needed for future missions.
In this paper, we carry out a quantitative investigation of this problem. 

To set meaningful detection goals, some target signals are needed, for which we consider two different scenarios for the search for \ac{BGRS}.
\begin{itemize}
\item The first is to check if there is any inconsistency in the \ac{GR} predicted signals. 
For this we consider two key predictions of \ac{GR} in the merger of \acp{MBHB}: the nonlinear $(2,2,0)\times(2,2,0)$ ringdown mode and the displacement memory.
These signals contain the less explored nonlinear effects of \ac{GR} and might have the opportunity to reveal the presence of \ac{BGRS}.
\item The second is to directly search for a putative \ac{BGRS}.
Unfortunately, so far no one is sure if or what \ac{BGRS} can be detected with near-term detectors \cite{Luo:2025ewp}.
So we will try to motivate for one that is logically as obvious as possible.
The \ac{BGRS} we consider is motivated by an emergent gravity picture, where gravity is believed to be the effect of a hidden fluid \cite{Mei:2022ksw,Mei:2022kcf}, and a black holes forms when the underlying hidden fluid is forced into an extremely high density phase.
We call this the \ac{FBH} model.
In analogy to the violent collision of normal matter systems, during the merger of massive \acp{FBH}, some mini black holes might be produced and, when the major remnant black hole is near extremal, may temporarily become satellite black holes orbiting the remnant black hole, forming \acp{EMRI}.
We call such systems \acp{iEMRI}, whose detection would provide smoking gun evidence for new physics beyond \ac{GR}. 
\end{itemize}
We set the sensitivity requirements to be such that, given a chosen detector and a target signal, the detector should be able to detect the target signal from a significant portion of the events predicted by a chosen \ac{MBHB} population model, such as Q3d, Q3nd and pop III \cite{Barausse:2012fy,Klein:2015hvg}.
In this way, one can hope to confirm/exclude with good confidence if there is any sign of \ac{BGRS} in the \ac{GW} signals. 

For reference detectors, we consider TianQin \cite{TianQin:2015yph,TianQin:2020hid,Luo:2025sos}, 
LISA \cite{LISA:2024hlh}, and $\mu$Ares \cite{Sesana:2019vho}.
These detectors are chosen because they represent space-based \ac{GW} missions with a similar design but a vastly different choice of arm-lengths.
Their sensitivity curves have a similar analytical form and are all controlled by two core noise parameters: 
the one way displacement measurement noise, $S_x^{1/2}$, for the inter-satellite laser interferometry, 
and the residual acceleration noise, $S_a^{1/2}$, of the test masses (i.e., inertial reference) along the sensitive directions. 
For each detector, we will keep its arm-length fixed and study how the noise parameters, $S_a^{1/2}$ and $S_x^{1/2}$, should be improved to achieve the desired sensitivity goals.
 In addition to the above, there are also other space-based \ac{GW} detectors, such as Taiji \cite{Luo:2021qji} and DECIGO \cite{Komori:2025pjx}. 
The results obtained in this study can also be of some reference value for them.

The most notable finding of the study is that the required improvements strongly depend on the actual population model of \acp{MBHB}, with Q3d always leading to the least stringent requirements while pop III always leads to the most demanding requirements.
Another notable feature is that, in the most demanding detection scenarios involving the putative \ac{iEMRI} signal or the pop III population model, the required improvements in $S_a^{1/2}$ and $S_x^{1/2}$ can go up to $4-9$ orders of magnitude, imposing a significant challenge for future development.

The paper is organized as follows.
In section \ref{sec:signal}, we introduce the main types of target signals and reference detectors to be used for this study. 
In section \ref{sec:rst}, we describe the methods and present the main results of the paper.
In section \ref{sec:sum}, we conclude with a brief summary.

\section{The target signals and reference detectors}
\label{sec:signal}

In this section, we briefly introduce the target signals that will be used for this study. 
These signals include the  $(2,2,0)\times(2,2,0)$ ringdown mode, the displacement memory, and \ac{iEMRI}. 
These signals are chosen because they are not only interesting targets for searching for new physics beyond \ac{GR}, but are also relatively difficult to detect.
We also introduce the three \ac{GW} detectors that are used as possible starting points for future improvement.

\subsection{The $(2,2,0)\times(2,2,0)$ ringdown mode}

After the merger of a \ac{MBHB}, the resulting compact remnant settles down to a stationary black hole by emitting \acp{GW}, which is the so called ringdown stage. 
The corresponding waveforms can be modeled by the sum of first-order \acp{QNM}, labeled by three indices $(\ell,m,n)$ \cite{Berti:2009kk}. 
When \ac{SNR} is large, nonlinear effects may also become important in the modeling of ringdown signals. 
The most relevant ones are the quadratic \acp{QNM}, which are solutions to the second-order perturbation theory \cite{Campanelli:1998jv} and are driven by quadratic combinations of the first-order \acp{QNM} \cite{Ioka2007}, and so the quadratic \acp{QNM} are labeled by the corresponding first-order \acp{QNM}, $(\ell_1,m_1,n_1)\times(\ell_2,m_2,n_2)$.
Understanding and detecting quadratic \acp{QNM} is not only an important direct test of the nonlinearities of \ac{GR}, but is also important for the better modeling of the early ringdown signals.

Just like the first-order \acp{QNM}, the second-order ones are also damped sinusoids,  whose oscillation frequencies, damping times, and amplitudes are related to the first-order ones through \cite{Nakano2007,Ioka2007, London2014},
\bea \omega_{(l_1,m_1,n_1)(l_2,m_2,n_2)}&=&\omega_{l_1,m_1,n_1} +\omega_{l_2,m_2,n_2}\,,\nn\\
\tau_{(l_1,m_1,n_1)(l_2,m_2,n_2)}^{-1}&=&\tau_{l_1,m_1,n_1}^{-1}+\tau_{l_2,m_2,n_2}^{-1}\,,\nn\\
A_{(l_1,m_1,n_1)(l_2,m_2,n_2)}&=&\mu_{(l_1,m_1,n_1)(l_2,m_2,n_2)}\nn\\
&&\times A_{l_1,m_1,n_1}A_{l_2,m_2,n_2}\,.\label{eq:srm}\eea
There are two methods to determine the ratio factor $\mu_{(l_1,m_1,n_1)(l_2,m_2,n_2)}$.
The first involves directly fitting Eq. \eqref{eq:srm} to existing numerical relativity data \cite{London2014,Mitman:2022qdl,Cheung:2022rbm}, and the second involves solving second-order perturbation equations using numerical methods \cite{Ma:2024qcv,Khera:2024bjs,Redondo-Yuste:2023seq,Zhu:2024rej}.
The results of both methods are consistent with each other, and it has been found that $\mu_{(l_1,m_1,n_1)(l_2,m_2,n_2)}$ only depends on the properties of the remnant black hole, especially its spin \cite{Khera:2024bjs,Redondo-Yuste:2023seq}. 
In this paper, we will focus on the detection of the $(2,2,0)\times(2,2,0)$ mode, for which \cite{Cheung:2022rbm}
\bea \mu_{(2,2,0)(2,2,0)}=0.1637\,.\eea

So far, the evidence for quadratic \acp{QNM} in \ac{GW} signals is still very limited.
In GW250114, the signal with the highest SNR (nearly 80) to date, clear evidence of higher-order ringdown modes and overtones has been established \cite{LIGOScientific:2025rid,LIGOScientific:2025wao}. 
It has been pointed out that including the second-order $(2,2,0)\times(2,2,0)$ mode can better explain the ringdown signal of GW250114, and $(2,2,0)\times(2,2,0)$ fits the data better than the $(4,4,0)$ mode \cite{Yang:2025ror}.
By including inspiral and merger data, a Bayes factor of 74 has been found to support the existence of the $(2,2,0)\times(2,2,0)$ mode in the ringdown data in \cite{Wang:2026rev}. 
For future space-based \ac{GW} detectors, TianQin and LISA can detect $2–15$ and $6-50$ events, respectively, that contain a detectable contribution from the $(2,2,0)\times(2,2,0)$ mode, with roughly half of which are resolvable despite mixing from other modes \cite{Shi:2024ttu}. 

\subsection{The displacement memory}

After the passage of \ac{GW}, the relative separation between two objects in spacetime experiences a permanent displacement, this effect is known as the \ac{GW} memory effect \cite{Zeldovich:1974gvh,Christodoulou:1991cr,Blanchet:1992br}. 
In an asymptotically flat spacetime, the memory effect has been shown to closely relate to various \ac{BMS} transformations \cite{Strominger:2014pwa,Pasterski:2015tva,Nichols:2018qac,Compere:2019gft}.
Meanwhile, the memory effect has been shown to be closely related to Weinberg’s soft graviton theorem \cite{Weinberg:1965nx}, which leads to the so-called infrared triangle, which connects the memory effect, asymptotic symmetries, and soft theorems \cite{Strominger:2017zoo}. 
In addition to \ac{GR}, memory effects have also been studied in other modified theories of gravity, such as Brans–Dicke theory \cite{Seraj:2021qja,Tahura:2021hbk}, Scalar-Tensor theory \cite{Du:2016hww}, Chern–Simons theory \cite{Hou:2021oxe,Hou:2021bxz} and Einstein-\AE ther theory \cite{Hou:2023pfz}.
Therefore, the detection of the \ac{GW} memory effect is of great importance, such as testing GR in the nonlinear regime \cite{Christodoulou:1991cr,Du:2016hww}, 
improving the precision of parameter estimation for \ac{GW} signals \cite{Ashtekar:2019viz,Sun:2022pvh}, 
and helping to study the asymptotic symmetry of spacetime \cite{Goncharov:2023woe}.

The \ac{BMS} symmetry allows a neat classification of the different types of memory effect: displacement memory, spin memory, and center-of-mass memory, which are related in one to one correspondence to supertranslation, superrotation, and superboost, respectively. 
Among the three, the displacement memory is always the strongest and will be chosen as a detection target for this paper.

The study of the detectability of the \ac{GW} memory effect started in the 1980s \cite{Braginsky:1987kwo}. 
After the detection of GW150914, existing ground-based detectors have been found to remain insufficient for direct observation of the memory effect, neither from a single event \cite{Lasky:2016knh} nor from the joint analysis of multiple events from GWTC-1, GWTC-2, and GWTC-3 \cite{Hubner:2019sly,Hubner:2021amk,Cheung:2024zow}. 
However, it has been found that next-generation detectors, such as the Cosmic Explorer \cite{LIGOScientific:2016wof} and the Einstein Telescope \cite{Punturo:2010zz}, may have the potential to directly detect the \ac{GW} memory effect\cite{Goncharov:2023woe}. 
The prospect of detecting the memory effect with a space-based \ac{GW} detector has been analyzed for TianQin in \cite{Sun:2022pvh,Sun:2024nut} and LISA in \cite{Islo:2019qht,Inchauspe:2024ibs}. 
Both detectors have the opportunity to detect some \ac{MBHB} signals that contain a detectable memory effect.
The possibility of detecting the memory effect using pulsar timing arrays has been explored in \cite{Seto:2009nv,vanHaasteren:2009fy,Pshirkov:2009ak,Cordes:2012zz,Madison:2014vca,NANOGrav:2015xuc}. 
Based on approximately 12.5 years of NANOGrav data, a Bayes factor of 2.8 has been obtained for the memory effect \cite{NANOGrav:2023vfo}.

\subsection{iEMRI}

There are largely two different ways to model and theorize gravity using only particles.\footnote{We do not consider theories that involve extended fundamental objects, such as strings or branes.}
The first is to assume that gravity is mediated by a fundamental force carrier, usually a boson called graviton.
In this case, gravity is a fundamental interaction of nature, and the spacetime metric is treated as the fundamental gravitational field, but the quantization of which has faced a lot of difficulties.
The second is to assume that gravity is caused by the coupling of matter to some hidden condensed matter system, usually made of interacting fermions.
In this case, gravity is emergent \cite{Barcelo:2001tb,Barcelo:2005fc,Hu:2009jd,Sindoni:2011ej,Carlip:2012wa,Linnemann:2017hdo}, and the spacetime metric is expected to be determined by either the properties of the hidden condensed matter system itself or its phonon-like excitations.
At the macroscopic scale, one expects that an interacting particle system can be treated as a fluid in the general sense, and so gravity is also expected to be determined by the properties of a fluid, which we call hidden fluid.

A quantitative relation between spacetime metric and hidden fluid has been explored in \cite{Mei:2022ksw,Mei:2022kcf} and some intriguing results have been found.
For example, the density of hidden fluid underlying a Schwarzschild black hole is given by
\bea \varrho(r)=\frac{\varrho_0}{1-2M/r}\,,\label{density}\eea
where $M$ is the mass of the black hole, $r$ is the Boyer-Lindquist radial coordinate, $\varrho_0$ is the average density of hidden fluid at spatial infinity.
The divergence of the density at the black hole horizon implies that some new physics must come into play before the horizon is reached.
If an analogy to normal matter can be made, then hidden fluid is expected to be forced into an incompressible phase, whose density cannot be further increased, when the density of hidden fluid reaches a high critical value $\varrho_{\rm cri}=\varrho_0/(1-2M/r_{\rm cri})$.
In this case, the density of hidden fluid is governed by \eqref{density} outside a Schwarzschild black hole, while that inside the black hole becomes a constant.

\begin{figure}[htbp!]
\centering\includegraphics[width=0.4\textwidth]{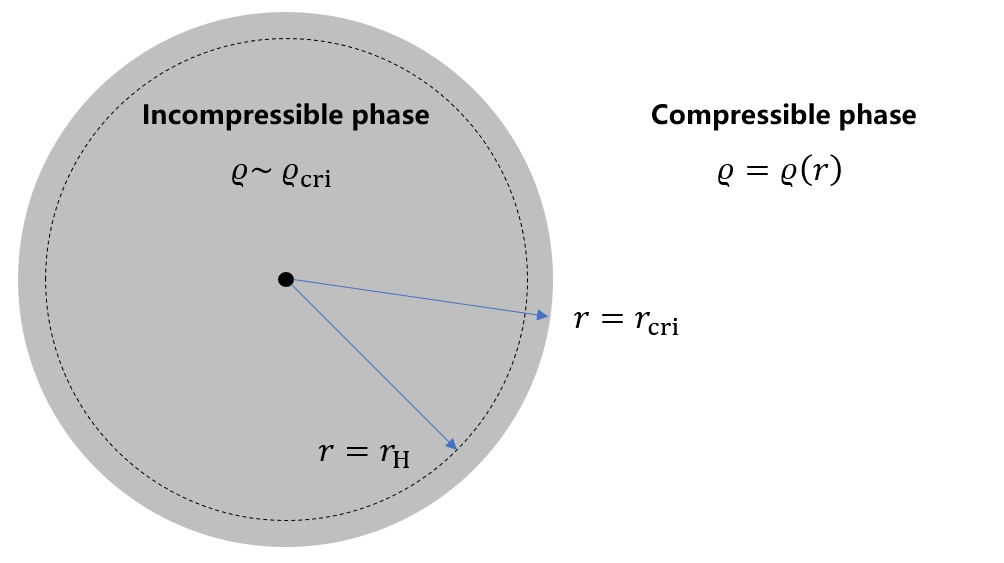}
\caption{An illustration of the \ac{FBH} model for a Schwarzschild black hole. }
\label{fig:model}
\end{figure}

Motivated by the above results, we consider the following \ac{FBH} model, illustrated for a Schwarzschild black hole in Fig. \ref{fig:model}.
In this model, the \ac{FBH} corresponds to a region of spacetime where hidden fluid is in the incompressible phase, with $\varrho\sim\varrho_{\rm cri}$.
The GR predicted event horizon is inside the \ac{FBH}, $r_{\rm H}<\varrho_{\rm cri}$, and is no longer relevant.
Outside the \ac{FBH}, hidden fluid is in the compressible phase and $\varrho(r)$ is a monotonically decreasing function of $r$, with $\varrho(r_{\rm cri})=\varrho_{\rm cri}$ and $\varrho(\infty)=\varrho_0$.
The exact detail of $\varrho(r)$ depends on the specific relation between spacetime metric and hidden fluid being explored, with (\ref{density}) being a specific example.

Due to the composite nature of \acp{FBH}, some behavior similar to normal matter systems might be expected.
For example, during the violent encounter of two \acp{FBH}, small segments of highly dense hidden fluid, which are effectively mini black holes, might be produced and ejected out of the collision region.
When the major remnant black hole is near extremal, there is the chance for an ejected mini black hole to get enough initial angular velocity to temporarily become a satellite black hole orbiting the remnant black hole. 
In this case, the pair becomes an \ac{iEMRI}.
The mini black holes are expected to fly only for a short time and then return to the major remnant black holes.
So, \acp{iEMRI} are expected to be short lived.

A detection of the \ac{iEMRI} signal will provide the smoking gun evidence for new physics beyond \ac{GR}.
This is because in \ac{GR} the second law of thermodynamics forbids a black hole from splitting into two or more black holes (see, e.g. \cite{Isi:2020tac} and references therein).
As a result, the close encounter of two black holes always produces only one black hole in \ac{GR}, even during a relativistic black hole collision \cite{Sperhake:2008ga,Sperhake:2009jz,Sperhake:2012me,Healy:2015mla,Bozzola:2022uqu,Healy:2022jbh,Healy:2024lhl,Fang:2025vnv}.

\subsection{The reference detectors}

We use the following three detectors as reference detectors for this study.
\begin{itemize}
\item TianQin consists of three geocentric satellites that form a nearly equilateral triangle with an arm length $L_{\mathrm{TQ}} \approx 1.7\times 10^8\,\mathrm{m}$. The sensitivity band of TianQin is assumed to be $10^{-4}-1$ Hz \cite{TianQin:2015yph,TianQin:2020hid,Luo:2025sos}.
\item LISA consists of three spacecraft that form a nearly equilateral triangle with an arm length $L_{\mathrm{LISA}}\approx 2.5 \times 10^9\,\mathrm{m}$. 
It will operate in an Earth-like heliocentric orbit, approximately $20^\circ$ apart from Earth. The LISA sensitivity band is assumed to be $2\times10^{-5}-1$ Hz \cite{LISA:2024hlh}.
\item $\mu$Ares is a future $\mu$-Hz space-based detector consists of two equilateral triangle constellations with an arm length of $L_{\mathrm{\mu}Ares}=4.3\times10^{11}\mathrm{m}$. The sensitivity band of $\mu$Ares is assumed to be $10^{-7}-1$ Hz \cite{Sesana:2019vho}.
In this study, we only consider one of the two constellations of $\mu$Ares.
\end{itemize}
These detectors are chosen because they have a good prospect of detecting \ac{MBHB} signals, their sensitivities share the same set of key noise parameters, and their arm lengths take a hierarchy of different values.
These are helpful for making a meaningful comparison.

The sky-averaged sensitivities of TianQin and LISA are given by:
\begin{widetext}
\bea S_n^{\mathrm{TQ}}(f)&=&\frac{10}{3L_{\mathrm{TQ}}^2}\Big[ S_x^{\mathrm{TQ}} + \frac{4S_a^{\mathrm{TQ}}}{(2\pi f)^4}\Big( 1 + \frac{10^{-4} \mathrm{Hz}}{f}\Big)\Big] \times \Big[ 1 + 0.6 \Big(\frac{f}{f_*^{\mathrm{TQ}}}\Big)^2\Big]\,,\label{eq:sensitivity_TQ}\\
S_n^{\mathrm{LISA}}(f)&=& \frac{10}{3L_{\mathrm{LISA}}^2} \Big\{ S_x^{\mathrm{LISA}}(f) + \frac{S_a^{\mathrm{LISA}}(f)}{(2\pi f)^4} \Big[ 1 + \cos^2 \Big( \frac{f}{f_*^{\mathrm{LISA}}} \Big) \Big] \Big\} \times \Big[ 1 + \frac{6}{10} \Big( \frac{f}{f_*^{\mathrm{LISA}}} \Big)^2 \Big]\,.\label{eq:sensitivity_LISA}\eea
\end{widetext}
Here, $f_*^{\mathrm{TQ}}=c/2\pi L_{\mathrm{TQ}}\approx0.28 \mathrm{Hz}$ is the transfer frequency of TianQin, 
$(S_x^{\mathrm{TQ}})^{1/2}=1\times 10^{-12}\mathrm{m/Hz^{1/2}}$ is the noise of the one-way displacement measurement, and 
$(S_a^{\mathrm{TQ}})^{1/2}=1\times10^{-15}\mathrm{m/s^2/Hz^{1/2}}$ is the residual acceleration noise \cite{TianQin:2015yph,TianQin:2020hid,Luo:2025sos}.
Similarly, $f_*^{\mathrm{LISA}} = 1/(2\pi L_{\mathrm{LISA}}) \approx 0.02\,\mathrm{Hz}$ is the transfer frequency of LISA, and
\bea S_x^{\mathrm{LISA}}(f)&=& S_x^{\mathrm{LISA}}\left[ 1 + \left( \frac{2 \times 10^{-3}\,\mathrm{Hz}}{f} \right)^4 \right]\,,\nn\\
S_a^{\mathrm{LISA}}(f)&=&S_a^{\mathrm{LISA}}\left[ 1 + \left( \frac{4 \times 10^{-4}\,\mathrm{Hz}}{f} \right)^2 \right] \nn\\ 
&&\qquad\times\left[ 1 + \left( \frac{f}{8 \times 10^{-3}\,\mathrm{Hz}} \right)^4 \right]\,,\label{eq:SaSx_LISA}\eea
where $(S_x^{\mathrm{LISA}})^{1/2}=1.5 \times 10^{-11}\mathrm{m/Hz}^{1/2}$ and 
$(S_a^{\mathrm{LISA}})^{1/2}=3\times10^{-15}\mathrm{m/s}^2/\mathrm{Hz}^{1/2}$ \cite{Robson:2018ifk,LISA:2024hlh}.

The analytical formula for the sensitivity of $\mu$Ares is currently unavailable. 
We assume that it has the same structure as \eqref{eq:sensitivity_LISA}, replacing LISA everywhere with $\mu$Ares, and fit it to Fig. 1 of \cite{Sesana:2019vho} to obtain the following:
\bea S_x^{\mu\mathrm{Ares}}(f) &=& S_x^{\mu\mathrm{Ares}} \left[ 1 + \left( \frac{1 \times 10^{-3}\,\mathrm{Hz}}{f} \right)^4 \right]\,,\nn\\
S_a^{\mu\mathrm{Ares}}(f) &=& S_a^{\mu\mathrm{Ares}} \left[ 1 + \left( \frac{4 \times 10^{-6}\,\mathrm{Hz}}{f} \right) \right] \nonumber\\ 
&&\qquad\times\left[ 1 + \left( \frac{f}{8 \times 10^{-3}\,\mathrm{Hz}} \right)^4 \right]\,,\eea
where $(S_x^{\mu\mathrm{Ares}})^{1/2}=3.5\times10^{-11}\mathrm{m/Hz}^{1/2}$ and 
$(S_a^{\mu\mathrm{Ares}})^{1/2}=2\times10^{-15}\mathrm{m/s}^2/\mathrm{Hz}^{1/2}$. 

A comparison of the sensitivities of the three reference detectors can be found in Fig.~\ref{fig:waveform_sensitivity}.

\section{Method and results}
\label{sec:rst}

In addition to the three types of target signals and three reference detectors, we also consider three population models of merging \acp{MBHB}: Q3d, Q3nd and pop III \cite{Barausse:2012fy,Klein:2015hvg}. 
For each population model, we have 10 independent samples, each containing a different number of sources.
We combine them to get a super-sample for each population model, containing 405, 6122 and 8735 sources in total for Q3d, Q3nd and pop III, respectively.
In this section, all the population-based plots are made using the super-samples.
So, if one wants to estimate some detection number by counting individual sources in a plot, one should divide the result by 10.

Each combination of a source, a target signal, and a reference detector is treated as a detection problem.
For each detection problem, we want to find the optimal values of $(S_a^{1/2}, S_x^{1/2})$, which not only need to allow the target signal to be detected with enough \ac{SNR} but also need to place the least demanding requirement on the development of technology.
To calculate \ac{SNR}, we use
\bea\rho=\sqrt{(h|h)}\,,\eea
where $h$ is the waveform of the target signal. 
The inner product is defined as:
\bea (a(f)|b(f)) = 2 \int_{f_{\mathrm{low}}}^{f_{\mathrm{high}}} \frac{a^*(f)b(f) + a(f)b^*(f)}{S_n(f)} df\,,\eea
where $S_n(f)$ is the sky-averaged sensitivity of the detector under consideration.
Throughout the study, we fix the detection threshold as $\rho_{\rm th}=8$.
This threshold has been used in some previous studies \cite{Shi:2024ttu,Sun:2022pvh}, but its exact value is not expected to bring an order of magnitude difference to the main conclusions of this paper.

\begin{figure}[htbp!]
\centering
\includegraphics[width=0.9\linewidth]{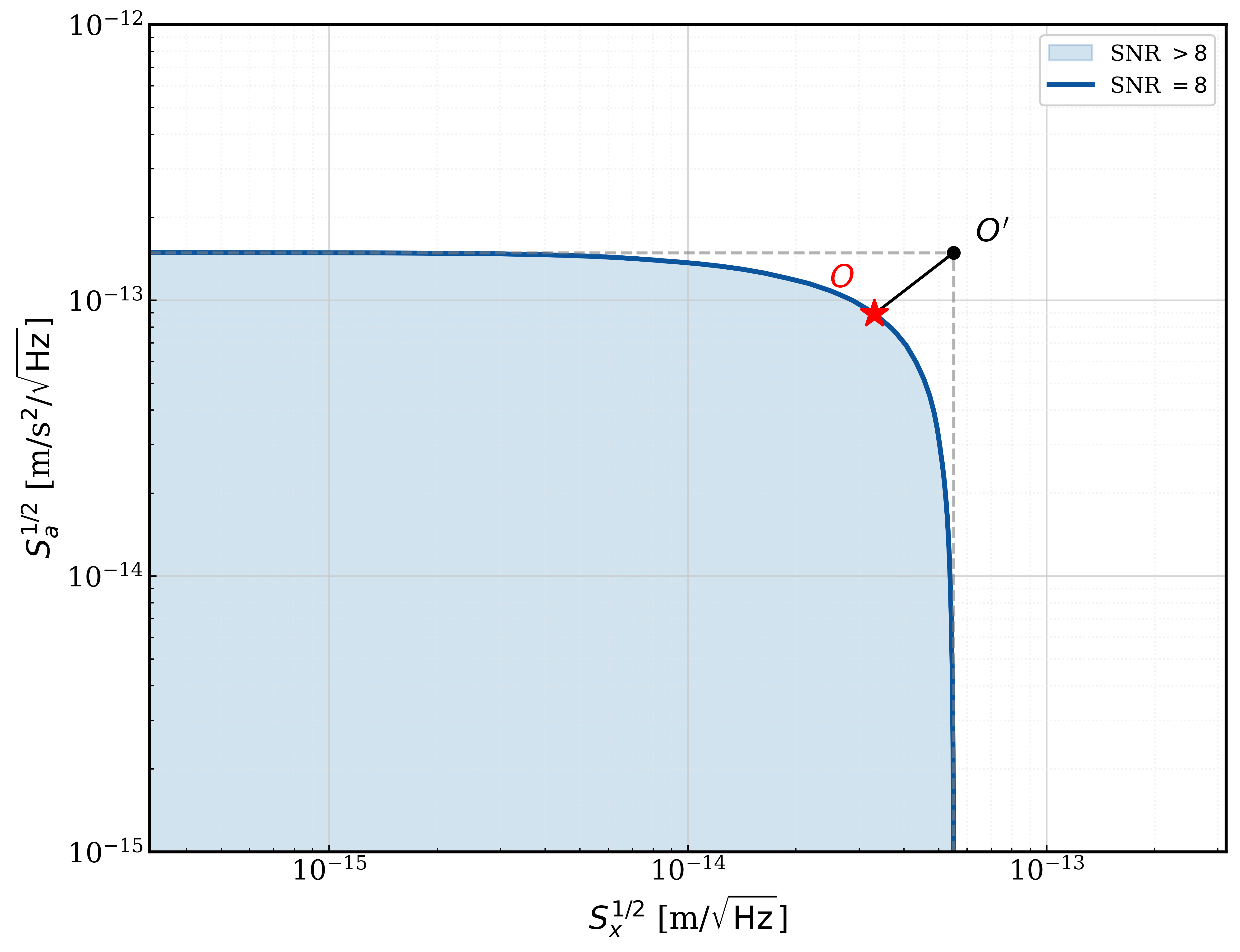}
\caption{The allowed values of $(S_a^{1/2}, S_x^{1/2})$ for a detection problem.
In the plot, the target signal is the $(2,2,0) \times (2,2,0)$ ringdown mode,
the reference detector is TianQin, 
and the injected source parameters are: 
$m_1 = 3\times 10^4 \mathrm{M_\odot}$, 
$m_2 = 2\times 10^4 \mathrm{M_\odot}$, 
$z=1$, 
$\iota = \pi/4$, and 
$\phi = 0$.
See main text for more explanation.}
\label{fig:optimal_SaSx}
\end{figure}

For each detection problem, \ac{SNR} becomes a function of the noise parameters $\rho=\rho(S_a^{1/2}, S_x^{1/2})$, and this function has a remarkable feature as illustrated in Fig. \ref{fig:optimal_SaSx}.
To understand the behavior, note $\rho(S_a^{1/2},0)=8$ and $\rho(0, S_x^{1/2})=8$ lead to constant solutions of $S_a^{1/2}$ and $S_x^{1/2}$, respectively, so the blue curve corresponding to $\rho(S_a^{1/2}, S_x^{1/2})=8$ is nearly horizontal when $S_x^{1/2}\rightarrow0$ and nearly vertical when $S_a^{1/2}\rightarrow0$.
Thus, a horizontal line and a vertical line (the dashed ones) can be drawn that are tangential to the blue curve and meet at the point $O'$.
In the region where the blue curve turns from horizontal to vertical, the contributions of $S_a^{1/2}$ and $S_x^{1/2}$ are comparable to each other.
This is where we pick the optimal values for $(S_a^{1/2}, S_x^{1/2})$ for each detection problem.
In practice, one would compare the technical difficulties in improving $S_a^{1/2}$ and $S_x^{1/2}$, and then decide which value to choose.
For this study, however, we will simply draw a line of slope 1\footnote{For this we always assume a log-log plot and take the units of axes to be same as in Fig. \ref{fig:optimal_SaSx}.} that starts from the point $O'$ and intersects the blue curve at the point $O$, and then take the latter to be the optimal choice of $(S_a^{1/2}, S_x^{1/2})$.

For a given target signal and a given reference detector, each source determines an optimal choice of $(S_a^{1/2}, S_x^{1/2})$, corresponding to a point in the parameter space of $(S_a^{1/2}, S_x^{1/2})$.
The results for all sources in a super-sample then determine a distribution.
The results are plotted in Figs.~\ref{fig:tianqin}-\ref{fig:muares}.
Some source selection has been performed when making the plots.
This is necessary because some of the target signals impose additional restrictions.
\begin{itemize}
\item \textbf{$(2,2,0) \times (2,2,0)$ mode}: All sources are used.
\item \textbf{Displacement memory}: Only sources with a mass ratio $q\in [1, 25]$ are used. 
Our calculation of the memory waveform depends on \texttt{IMRPhenomXHM} \cite{Garcia-Quiros:2020qpx,Pratten:2020fqn,Colleoni:2020tgc}, which has been calibrated with numerical relativity within $q \in [1, 18]$ but is believed to have a reasonable accuracy outside of this range \cite{Garcia-Quiros:2020qpx}.
To transform the memory waveform from the time domain to the frequency domain, we use the method of \cite{Valencia:2024zhi}.
\item \textbf{iEMRI}: We only consider sources in which the remnant black hole is near extremal, and only consider the radiation from the last cycle before plunge.
The corresponding waveform is generated with \texttt{kerrgeodesic\_gw} \cite{Gourgoulhon:2019iyu}, which supports spins $0.9$ and $0.95$ for the central black hole.
For these reasons, we only consider sources with final spins in the range $[0.875, 0.975]$.
Those in the range $[0.875, 0.925]$ are approximated to have spin $0.9$, and those in the range $[0.925, 0.975]$ are approximated to have spin $0.95$. 
We also assume that the mass of the mini black hole is always $10^{-6}$ times that of the central black hole.
\end{itemize}
For each distribution, we define its center to be at the center-of-mass location, assuming log-log scale and equal weight for all the sources. 
See the last rows of Figs.~\ref{fig:tianqin}-\ref{fig:muares}.
Given a target signal, a reference detector and a population model, the required improvement in technology is characterized by the difference between the current location of the detector in the parameter space of $(S_a^{1/2}, S_x^{1/2})$ and the center of the corresponding distribution.

\begin{figure*}[htbp!]
\centering
\includegraphics[width=0.9\linewidth]{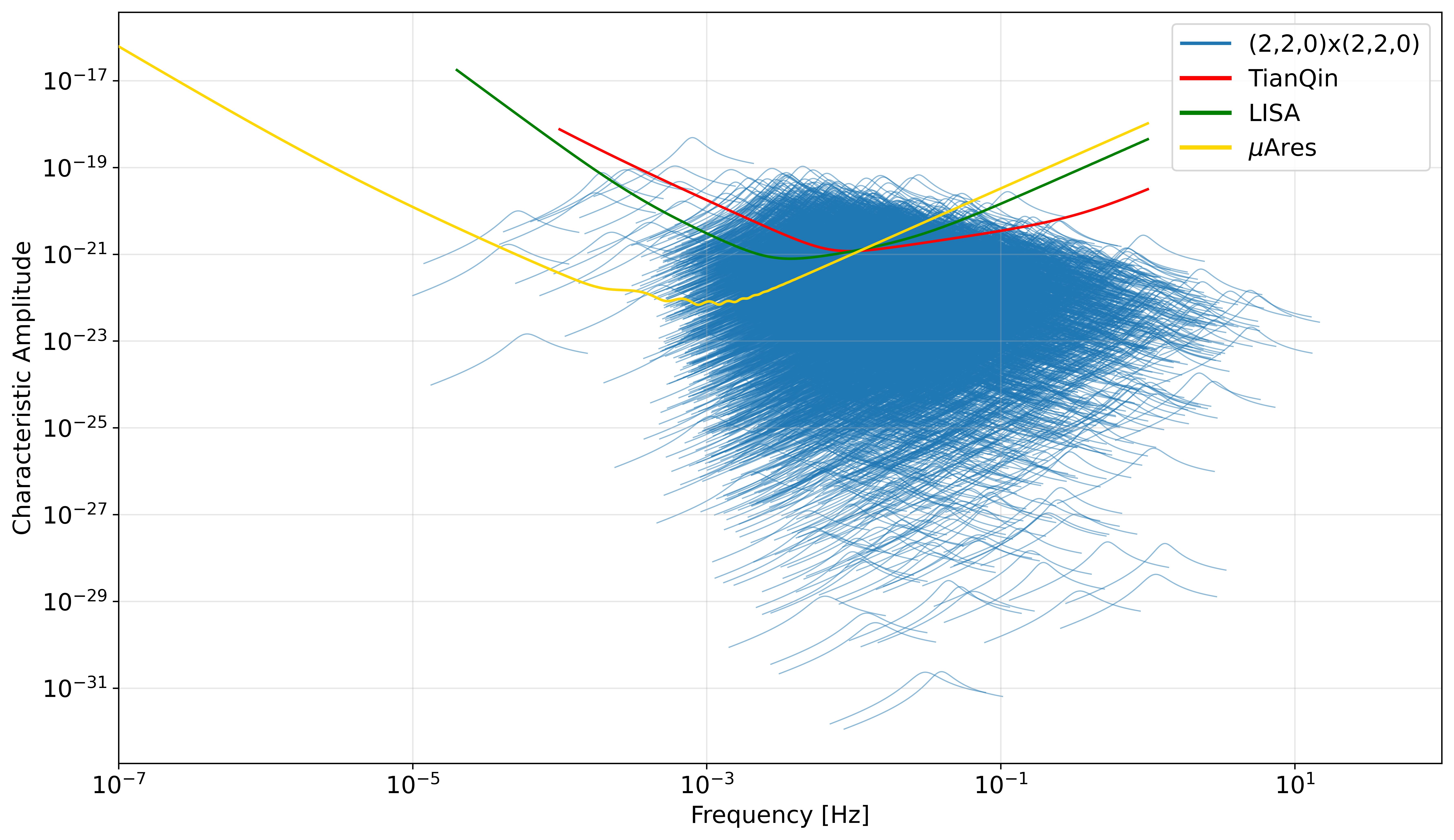}
\caption{Signal of the $(2,2,0) \times (2,2,0)$ ringdown mode for all sources in the Q3nd super-sample. 
The sensitivity curves of the three reference detectors are also shown as a comparison.}
\label{fig:waveform_sensitivity}
\end{figure*}

Fig.~\ref{fig:tianqin} contains the results for TianQin.
There are a few obvious features.
\begin{itemize}
\item To detect the $(2,2,0) \times (2,2,0)$ ringdown mode, it will be more effective if $S_x$ can be significantly improved. 
The improvement in $S_x^{1/2}$ should be about 2, 3 and 6 orders of magnitude if the population model is Q3d, Q3nd and pop III, respectively.
\item To detect displacement memory, it will be more effective if $S_a$ can be significantly improved.
The improvement in $S_a^{1/2}$ should be about 2, 3 and 4 orders of magnitude if the population model is Q3d, Q3nd and pop III, respectively.
\item To detect \ac{iEMRI}, both $S_a$ and $S_x$ need to be significantly improved.
The improvement in $S_a^{1/2}$ should be about 4 orders of magnitude for all population models, while the improvement in $S_x^{1/2}$ should be about 4, 5 and 7 orders of magnitude if the population model is Q3d, Q3nd and pop III, respectively.
\item The requirements from pop III are always the most stringent, typically of 3-7 orders of magnitude, while those from Q3d are always the least demanding, typically of 0-4 orders of magnitude.
\end{itemize}

In the detection of the $(2,2,0) \times (2,2,0)$ ringdown mode, some of the distributions show clear straight edges. 
In particular, the distribution of Q3nd (first plot in the second row of Fig.~\ref{fig:tianqin}) is bounded both from the upper-left edge and from the lower-right edge.
To understand this, in Fig.~\ref{fig:waveform_sensitivity} we plot all the waveforms of the $(2,2,0) \times (2,2,0)$ ringdown mode for all sources in the Q3nd super-sample. 
The waveforms of the lightest sources are partially outside the sensitivity band of TianQin, which is $10^{-4}-1$ Hz.
The part of their waveforms remaining in the band become nearly identical to each other, differing only by a scaling of their magnitudes.
This causes them to be distributed in the same line in the parameter space of $(S_a^{1/2}, S_x^{1/2})$, forming the upper-left edge.
For the heavy sources, one can see in Fig.~\ref{fig:waveform_sensitivity} that there is a significant amount of low frequency signals distributed near $10^{-3}$ Hz. 
These signals also have nearly identical waveforms that differ only by scaling of their magnitudes.
This also causes them to be distributed along the same line, forming the lower-right edge.

Compared to Fig.~\ref{fig:tianqin}, the most significant feature of Fig.~\ref{fig:lisa} and Fig.~\ref{fig:muares}, corresponding to LISA and $\mu$Ares, respectively, is that most of the sources are distributed very close to a narrow line.
To understand this, note that most signals in Fig.~\ref{fig:waveform_sensitivity} are distributed in frequencies higher than $10^{-3}$ Hz.
For the majority of them, \eqref{eq:SaSx_LISA} can be approximated as
\bea S_x^{\mathrm{LISA}}(f)&\approx& S_x^{\mathrm{LISA}}\,,\nn\\
S_a^{\mathrm{LISA}}(f)&\approx&S_a^{\mathrm{LISA}}\left(\frac{f}{8 \times 10^{-3}\,\mathrm{Hz}}\right)^4\,. \label{eq:SaSx_LISA2}\eea
Consequently, the most significant piece of frequency dependence cancels out in \eqref{eq:sensitivity_LISA} through the combination $S_a^{\mathrm{LISA}}(f)/(2\pi f)^4$, and $S_x$ and $S_a$ contribute to \eqref{eq:sensitivity_LISA} through a nearly frequency-independent combination, resulting in a constant ratio between their optimal values. 
Only for a small number of very massive sources, which have a significant contribution from frequencies near $10^{-3}$ Hz, can one observe some deviation from the narrow line distribution. 
The same is true for $\mu$Ares.

Another obvious feature is that, just like for TianQin, Q3d always places the least stringent requirements on technology improvement, typically no higher than $4$ orders of magnitude, while those from pop III are typically $3-6$ orders more demanding.

Some of the numerical results are summarized in Tables \ref{tab:tianqin}-\ref{tab:muares}, where  220Q stands for $(2,2,0) \times (2,2,0)$ and we have defined
\bea q_a=\frac{S_a^{\mathrm{opt}}}{S_a^{\mathrm{det}}}\,,\quad 
q_x=\frac{S_x^{\mathrm{opt}}}{S_x^{\mathrm{det}}}\,.\eea
Here, $S_a^{\mathrm{opt}}$ and $S_x^{\mathrm{opt}}$ represent the center values of a given distribution, and $S_a^{\mathrm{det}}$ and $S_x^{\mathrm{det}}$ represent the current values of the corresponding detector.

\begin{table}[htbp!]
\centering
\caption{The required improvement in the key noise parameters of TianQin}
\label{tab:tianqin}
\renewcommand{\arraystretch}{1.5}
\begin{tabular}{|c|c|c|c|c|}
\hline
\multicolumn{2}{|c|}{TianQin}   & \textbf{Q3d} & \textbf{Q3nd} & \textbf{pop III} \\
\hline
\multirow{2}{*}{220Q} 
& $q_a$ & 3.05                  & $3.08\times 10^{-1}$  & $8.45\times 10^{-3}$ \\
& $q_x$ & $1.75\times 10^{-2}$  & $1.25\times 10^{-3}$  & $1.40\times 10^{-6}$ \\
\hline
\multirow{2}{*}{Memory}
& $q_a$ & $2.71\times 10^{-2}$  & $4.94\times 10^{-3}$  & $1.81\times 10^{-4}$ \\
& $q_x$ & 5.69                  & $4.48\times 10^{-1}$  & $8.93\times 10^{-4}$ \\
\hline
\multirow{2}{*}{iEMRI}
& $q_a$ & $5.01\times 10^{-4}$  & $3.38\times 10^{-4}$  & $6.04\times 10^{-4}$ \\
& $q_x$ & $2.69\times 10^{-4}$  & $6.23\times 10^{-5}$  & $1.04\times 10^{-7}$ \\
\hline
\end{tabular}
\end{table}

\begin{table}[htbp!]
\centering
\caption{The required improvement in the key noise parameters of LISA}
\label{tab:lisa}
\renewcommand{\arraystretch}{1.5}
\begin{tabular}{|c|c|c|c|c|}
\hline
\multicolumn{2}{|c|}{LISA} & \textbf{Q3d} & \textbf{Q3nd} & \textbf{pop III} \\ 
\hline
\multirow{2}{*}{220Q}
& $q_a$ & $4.78\times10^{-1}$ & $2.65\times10^{-2}$ & $4.58\times10^{-6}$ \\ 
& $q_x$ & $6.63\times10^{-2}$ & $3.19\times 10^{-3}$ & $4.64\times 10^{-7}$ \\ 
\hline
\multirow{2}{*}{Memory}
& $q_a$ & $3.21\times10^{-1}$ & $6.59\times10^{-2}$ & $1.46\times10^{-3}$ \\ 
& $q_x$ & $4.81\times10^{-1}$ & $7.06\times10^{-2}$ & $6.25\times10^{-4}$ \\ 
\hline
\multirow{2}{*}{iEMRI}
& $q_a$ & $9.93\times10^{-4}$ &$ 3.08\times10^{-4}$ & $1.09\times10^{-7}$ \\ 
& $q_x$ & $2.37\times10^{-4}$ & $4.92\times10^{-5}$ & $1.05\times10^{-8}$ \\ 
\hline
\end{tabular}
\end{table}

\begin{table}[htbp!]
\centering
\caption{The required improvement in the key noise parameters of $\mu$Ares}
\label{tab:muares}
\renewcommand{\arraystretch}{1.5}
\begin{tabular}{|c|c|c|c|c|}
\hline
\multicolumn{2}{|c|}{$\mu$Ares} & \textbf{Q3d} & \textbf{Q3nd} & \textbf{pop III} \\ 
\hline
\multirow{2}{*}{220Q}
& $q_a$ & 1.54 & $6.56\times 10^{-2}$ & $7.17\times 10^{-6}$ \\ 
& $q_x$ & $9.32\times 10^{-2}$ & $3.68\times 10^{-3}$ & $5.30\times 10^{-7}$ \\ 
\hline
\multirow{2}{*}{Memory}
& $q_a$ & $19.6$ & 2.32 & $1.68\times 10^{-2}$ \\ 
& $q_x$ & $27.1$ & 2.84 & $1.51\times 10^{-2}$ \\ 
\hline
\multirow{2}{*}{iEMRI}
& $q_a$ & $5.97\times 10^{-3}$ & $1.22\times 10^{-3}$ & $1.68\times 10^{-7}$ \\ 
& $q_x$ & $5.08\times 10^{-4}$ & $6.41\times 10^{-5}$ & $4.63\times 10^{-9}$ \\ 
\hline
\end{tabular}
\end{table}

\section{Summary}
\label{sec:sum}

In this paper, we investigate how much improvement in sensitivity will be needed to search for a possible \ac{BGRS} with future space-based \ac{GW} detectors. 
The main purpose is to obtain a quantitative understanding of 
the prospect of searching for \ac{BGRS} simply by improving the sensitivity of a detector.

We use TianQin, LISA and $\mu$Ares as reference detectors and study how their key noise parameters, $S_a$ and $S_x$, should be improved to meet a set of detection goals, for which we use the $(2,2,0) \times (2,2,0)$ ringdown mode, displacement memory and iEMRI as target signals, and use Q3d, Q3nd and pop III as the \ac{MBHB} population models.
For a given type of target signal, a reference detector, and a population model, we determine how the noise parameters should be improved so that the target signal can be detected from a significant portion of the sources predicted by the population model.
In this way, one can hope to confirm/exclude with good confidence if there is any sign of \ac{BGRS} in the \ac{GW} signals. 

The most notable finding of our study is that the required improvement is strongly dependent on the population model of the \ac{GW} sources.
If \acp{MBHB} obey the Q3d population model, then no more than a 4 order of magnitude improvement in $(S_a^{1/2}, S_x^{1/2})$ will be needed for all three detectors, even for the most demanding \ac{iEMRI} signals.
In fact, in this case the current design of $\mu$Ares already has a very good ability to detect displacement memory, as is obvious from Table \ref{tab:muares}.
If, instead, \acp{MBHB} obey the pop III population model, then to detect the \ac{iEMRI} signals, about (4, 7), (7, 8) and (7, 9) orders of magnitude improvement in $(S_a^{1/2}, S_x^{1/2})$ will be needed for TianQin, LISA and $\mu$Ares, respectively.
In this case, $\mu$Ares will need about 2 orders of magnitude improvement in $(S_a^{1/2}, S_x^{1/2})$ to detect displacement memory.

It is also notable that, in the most demanding detection scenarios (those involving \ac{iEMRI} or pop III), the required improvements in $S_a^{1/2}$ and $S_x^{1/2}$ can go up to $4-9$ orders of magnitude.
With such a level of technological improvement, environmental disturbances and strong coexisting \ac{GW} signals can also become an obstacle.
For example, a space magnetic field can cause an acceleration noise of the order $(S_a^{\rm Mag})^{1/2}\sim 10^{-17}$ m/s$^2$/Hz$^{1/2}$ at around $10^{-3}$ Hz for inertial references \cite{Jia-Hui:2025hkv,Su:2025afa}, and a rough estimate shows that for a full \ac{MBHB} signal, an \ac{iEMRI} signal will only cause a mismatch in roughly the order $10^{-11}$.
Both problems appear to be very difficult to overcome.
As such, a significant challenge for future development can be expected if one needs to use the most demanding detection scenario to search for a \ac{BGRS}.

\section*{Acknowledgments}

The work has been supported in part by the National Key Research and Development Program of China (Grant No. 2023YFC2206700), and the Fundamental Research Funds for the Central Universities, Sun Yat-sen University.

\bibliographystyle{apsrev4-1}
\bibliography{ref}
\begin{figure*}[htbp!] 
\centering
\includegraphics[width=1\linewidth]{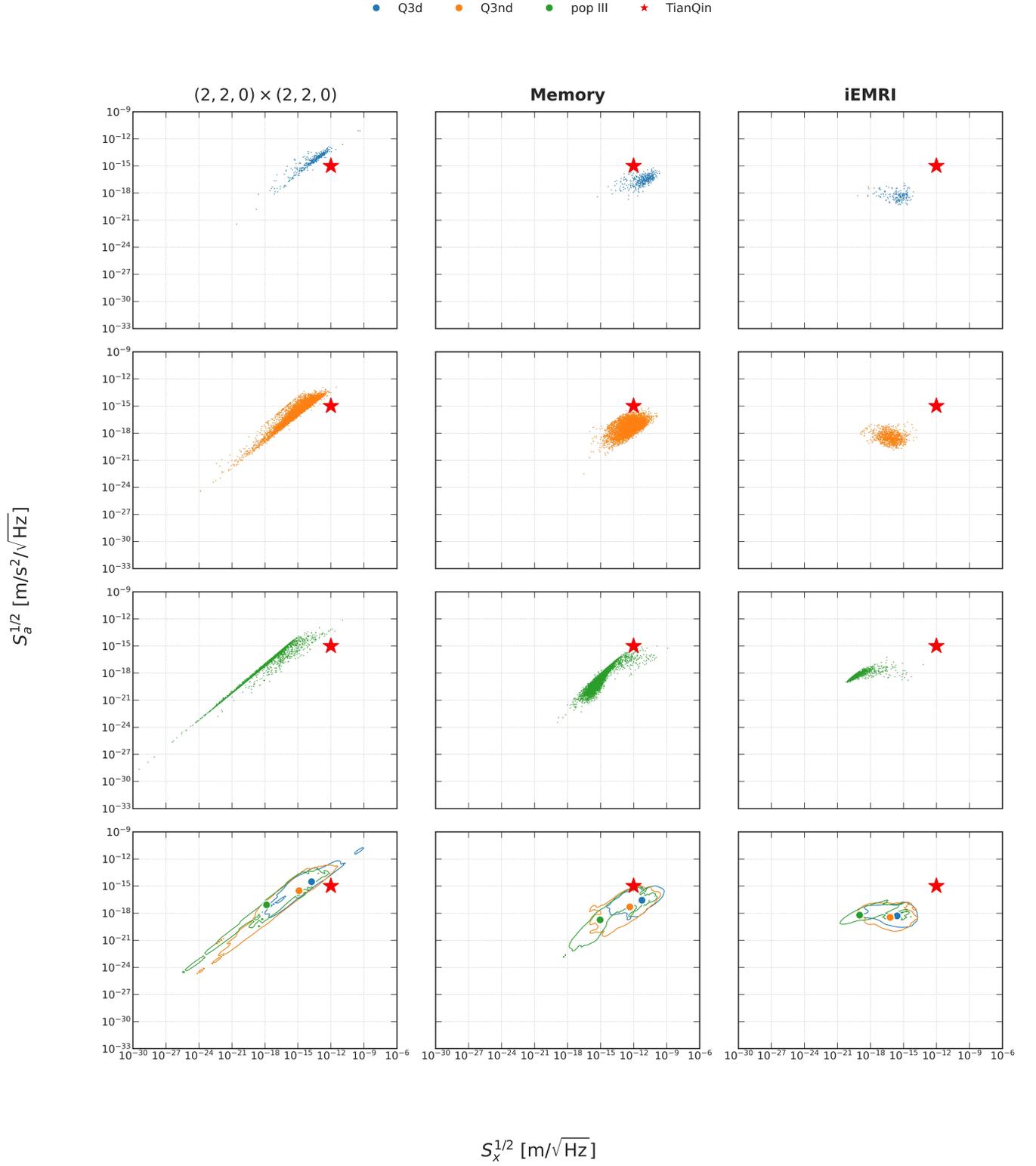}
\caption{Distribution of optimal values of $(S_a^{1/2}, S_x^{1/2})$ for TianQin.
The last row compares the major distribution regions and their centers for different population models.
See main text for more explanation.}
\label{fig:tianqin}
\end{figure*}

\begin{figure*}[htbp!]
\centering
\includegraphics[width=1\linewidth]{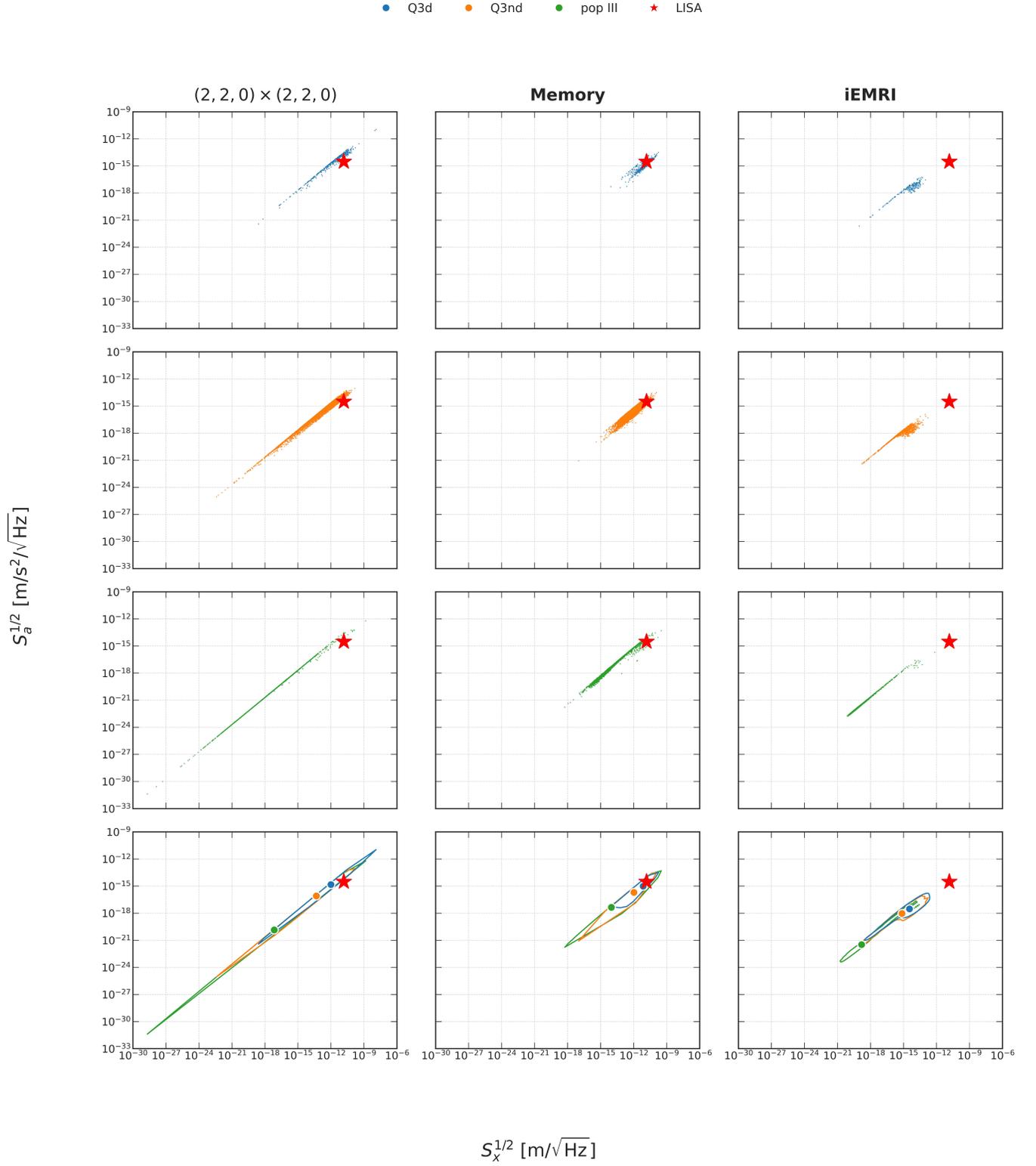}
\caption{Distribution of optimal values of $(S_a^{1/2}, S_x^{1/2})$ for LISA.
The last row compares the major distribution regions and their centers for different population models.
See main text for more explanation.}
\label{fig:lisa}
\end{figure*}

\begin{figure*}[htbp!]
\centering
\includegraphics[width=1\linewidth]{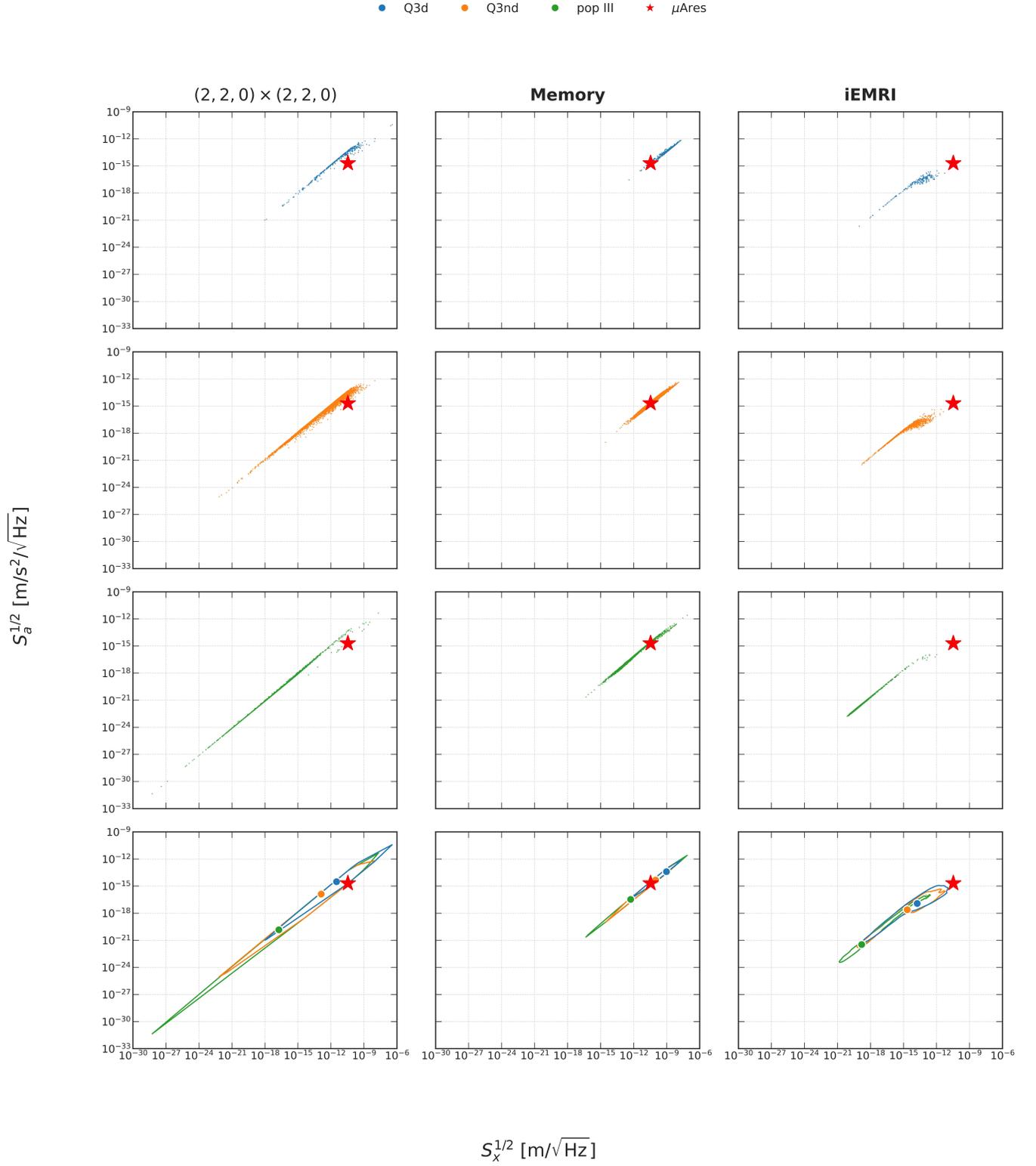}
\caption{Distribution of optimal values of $(S_a^{1/2}, S_x^{1/2})$ for $\mu$Ares.
The last row compares the major distribution regions and their centers for different population models.
See main text for more explanation.}
\label{fig:muares}
\end{figure*}
\end{document}